\documentclass[a4paper,11pt]{article}

\usepackage[dvips]{graphicx}
\usepackage{subfigure}
\usepackage{latexsym}
\usepackage{moreverb}
\usepackage{amssymb}
\usepackage{amsmath}
\usepackage{alltt}
\usepackage{makeidx}
\usepackage{ifthen}
\usepackage{pstricks}
\usepackage{eufrak}
\usepackage{float}
\usepackage{framed}
\usepackage{qip}
\usepackage{bbm}  
\usepackage{amsbsy}   
\usepackage{jeffe}

\newcounter{itm}
\newenvironment{myprotocol}[1]
  {\begin{minipage}{\columnwidth} 
    \begin{framed}\hspace{0ex} 
     \begin{minipage}{0.99\columnwidth}
       {\bf #1:}
       \setcounter{itm}{1}
       \begin{list}{\arabic{itm}.}{\usecounter{itm}}}
   {    \end{list}
       \vspace{-1.5ex} 
       \end{minipage} 
     \end{framed} 
    \end{minipage}\vspace{-0.6ex}}

\newenvironment{myfigure}[1]    
         {\begin{figure}[#1] \centering}
         { \end{figure}}

\allowdisplaybreaks
\DeclareMathSymbol{\PP}{\mathalpha}{AMSb}{"50}
\DeclareMathSymbol{\QQ}{\mathalpha}{AMSb}{"51}
\DeclareMathSymbol{\MM}{\mathalpha}{AMSb}{"42}
\DeclareMathSymbol{\NN}{\mathalpha}{AMSb}{"4E}
\DeclareMathSymbol{\ZZ}{\mathalpha}{AMSb}{"5A}
\DeclareMathSymbol{\RR}{\mathalpha}{AMSb}{"52}

\newcommand{\A}{\ensuremath{{\sf S}}} 
\newcommand{\B}{\ensuremath{{\sf R}}} 
\newcommand{\dA}{\ensuremath{\tilde{\sf S}}} 
\newcommand{\dB}{\ensuremath{\tilde{\sf R}}} 
\newcommand{\V}{\ensuremath{{\sf V}}} 
\renewcommand{\C}{\ensuremath{{\sf C}}} 
\newcommand{\dC}{\ensuremath{\tilde{\sf C}}} 

\newcommand{\qot}{{\sc qot}}
\newcommand{\epr}{{\sc epr}}
\newcommand{\eprqot}{\epr-\qot}
\newcommand{\comm}{{\sc comm}}
\newcommand{\eprcomm}{{\sc epr-comm}}
\newcommand{\BBqot}{{\sc bb84-qot}}
\newcommand{\BBeprqot}{{\sc bb84-epr-qot}}

\newcommand{\univ}{two-universal}
\newcommand{\gu}[1]{\ensuremath{\mathrm{H}_{#1}}}

\newcommand{\perm}[1]{\ensuremath{\Pi_n}}

\newcommand{\ol}[1]{\overline{#1}}

\newcommand{\assign}{\ensuremath{\kern.5ex\raisebox{.1ex}{\mbox{\rm:}}\kern -.3em =}}

\newcommand{\Qp}{Q^{+}}
\newcommand{\Qt}{Q^{\times}}
\newcommand{\qp}{q^{+}}
\newcommand{\qt}{q^{\times}}

\newcommand{\negl}[1]{\mathit{negl}({#1})}

\newcommand{\rs}{\ensuremath{\boldsymbol{\rho}}}

\newcommand{\nbit}{\{ 0,1 \} ^n}

\newcommand{\QtSt}{\Qt(S^{\times})}

\newcommand{\bg}[1]{\hat{#1}}
\newcommand{\ball}[1]{{B}^{#1}}

\newcommand{\Ba}[1]{\ensuremath{\mathcal{B}^{#1}}}
\renewcommand{\I}{\mathbbm{1}}

\newtheorem{theorem}{Theorem}[section]
\newtheorem{definition}[theorem]{Definition}
\newtheorem{lemma}[theorem]{Lemma}
\newtheorem{proposition}[theorem]{Proposition}
\newtheorem{corollary}[theorem]{Corollary}

\newcommand{\hf}{f}
\newcommand{\unif}{\mbox{\sc unif}}
\renewcommand{\Set}[2]{\{#1:\,#2\}}

\newcommand{\delete}[1]{}

\newcommand{\remove}[1]{}

\allowdisplaybreaks

\title{Cryptography in the \\Bounded Quantum-Storage Model\thanks{A
    preliminary version of this
    paper appeared in the proceedings of FOCS 2005 \cite{DFSS05}.}}

\newcounter{BRICS}
\newcounter{FICS}
\newcounter{SECOQC}
\newcounter{PROSECCO}

\newcommand{\ivan}{Ivan B. Damg{\aa}rd\thanks{Basic Research in Computer
      Science (BRICS),
      funded by the Danish National Research Foundation,
      Department of Computer Science, University of {\AA}rhus, {\tt\{ivan|salvail|chris\}@brics.dk}.
      }\setcounter{BRICS}{\value{footnote}}
     \thanks{FICS,
     Foundations in Cryptography and Security,
     funded by the Danish Natural Sciences Research Council.}\setcounter{FICS}{\value{footnote}}
 }

\newcommand{\louis}{Louis Salvail\footnotemark[\value{BRICS}]
                                 \footnotemark[\value{FICS}]
                                 \thanks{Supported in part by the
                                  European project PROSECCO.}
                                 \setcounter{PROSECCO}{\value{footnote}}
}

\newcommand{\christian}{Christian Schaffner\footnotemark[\value{BRICS}]
                                 \thanks{Supported by the
                                     European project SECOQC.}
                                 \setcounter{SECOQC}{\value{footnote}}
}

\newcommand{\serge}{Serge Fehr\thanks{Center for Mathematics and Computer Science (CWI),
    Amsterdam, Netherlands, {\tt fehr@cwi.nl}}
}
\author{\ivan \and \serge \and \louis \and \christian}

\begin{document}

\maketitle

\begin{abstract}
  We initiate the study of two-party cryptographic primitives with
  unconditional security, assuming that the adversary's {\em quantum}
  memory is of bounded size. We show that oblivious transfer and bit
  commitment can be implemented in this model using protocols where
  honest parties need no quantum memory, whereas an adversarial player
  needs quantum memory of size at least $n/2$ in order to break the
  protocol, where $n$ is the number of qubits transmitted. This is in
  sharp contrast to the classical bounded-memory model, where we can
  only tolerate adversaries with memory of size quadratic in honest
  players' memory size. Our protocols are efficient, non-interactive
  and can be implemented using today's technology. On the technical
  side, a new entropic uncertainty relation involving min-entropy is
  established.
\end{abstract}

\section{Introduction}
It is well known that non-trivial 2-party cryptographic primitives
cannot be securely implemented if only error-free communication is
available and there is no limitation assumed on the computing power
and memory of the players. Fundamental examples of such primitives are
bit commitment (BC) and oblivious transfer (OT). In BC, a committer $\C$
commits himself to a choice of a bit $b$ by exchanging information
with a verifier $\V$. We want that $\V$ does not learn $b$ (we say the
commitment is hiding), yet $\C$ can later chose to reveal $b$ in a
convincing way, i.e., only the value fixed at commitment time will be
accepted by $\V$ (we say the commitment is binding). In (Rabin) OT, 
a sender $\A$
sends a bit $b$ to a receiver $\B$ by executing some protocol in such a way that
$\B$ receives $b$ with probability $\frac12$ and nothing with probability
$\frac12$, yet $\A$ does not learn what was received.

Informally, BC is not possible with unconditional security since
hiding means that when 0 is committed, exactly the same information
exchange could have happened when committing to a 1. Hence, even if 0
was actually committed to, $\C$ could always compute a complete view of
the protocol consistent with having committed to 1, and pretend that
this was what he had in mind originally. A similar type of argument
shows that OT is also impossible in this setting.

One might hope that allowing the protocol to make use of quantum
communication would make a difference. Here, information is stored in
qubits, i.e., in the state of two-level quantum mechanical systems,
such as the polarization state of a single photon. It is well known
that quantum information behaves in a way that is fundamentally
different from classical information, enabling, for instance,
unconditionally secure key exchange between two honest players.
However, in the case of two mutually distrusting parties, we are not
so fortunate: even with quantum communication, unconditionally secure
BC and OT remain impossible~\cite{LC97,Mayers97}.

There are, however, several scenarios where these impossibility
results do not apply, namely:
\begin{itemize}
\item if the computing power of players is
  bounded,
\item if the communication is noisy,
\item if the adversary is under some physical limitation, e.g., the
  size of the available memory is bounded.
\end{itemize}

The first scenario is the basis of many well known solutions based on
plausible but unproven complexity assumptions, such as hardness of
factoring or discrete logarithms. The second scenario has been used to
construct both BC and OT protocols in various models for the
noise \cite{CK88,DFMS04,DKS99}. The third scenario is our focus here. In this
model, OT and BC can be done using classical communication assuming,
however, quite restrictive bounds on the adversary's memory
size \cite{CCM98,DHRS04}, namely it can be at most quadratic in the memory
size of honest players. Such an assumption is on the edge of being
realistic, it would clearly be more satisfactory to have a larger
separation between the memory size of honest players and that of the
adversary. However, this was shown to be impossible \cite{DM04}.

In this paper, we study for the first time what happens if instead we
consider protocols where quantum communication is used and we place a
bound on the adversary's {\em quantum} memory size.  There are two
reasons why this may be a good idea: first, if we do not bound the
classical memory size, we avoid the impossibility result of \cite{DM04}.
Second, the adversary's typical goal is to obtain a certain piece of
classical information that we want to keep hidden from him. However, 
if he cannot store all the quantum information that is sent, he
must convert some of it to
classical information by measuring. This may irreversibly destroy information, 
and we may be able to arrange it such that the adversary cannot afford to
lose information this way, while honest players can. 

It turns out that this is indeed possible: we present protocols for
both BC and OT in which $n$ qubits are transmitted, where honest
players need {\em no quantum memory}, but where the adversary must
store at least $n/2$ qubits to break the protocol. We emphasize that
no bound is assumed on the adversary's computing power, nor on his
classical memory. This is clearly much more satisfactory than the
classical case, not only from a theoretical point of view, but also in
practice: while sending qubits and measuring them immediately as they
arrive is well within reach of current technology, storing even a
single qubit for more than a fraction of a second is a formidable
technological challenge. Furthermore, we show that our protocols also
work in a non-ideal setting where we allow the quantum source to be
imperfect and the quantum communication to be noisy.

We emphasize that what makes OT and BC possible in our model
is not so much the memory bound per se, rather it is the loss
of information it implies on the part of the adversary.
Indeed, our results also hold if the adversary's memory device holds
an arbitrary number of qubits, but is imperfect is certain ways.
This is discussed in more detail in Section~\ref{noisymem}. 

Our protocols are non-interactive, only one party sends information
when doing OT, commitment or opening.  Furthermore, the commitment
protocol has the interesting property that the only message
is sent to the committer, i.e., it is possible to commit
while only {\em receiving} information.  Such a scheme clearly does
not exist without a bound on the committer's memory, even under
computational assumptions and using quantum communication: a corrupt
committer could always store (possibly quantumly) all the information
sent, until opening time, and only then follow the honest
committer's algorithm to figure out what should be sent to
convincingly open a 0 or a~1.  Note that in the classical 
bounded-storage model, it is known how to do time-stamping that is
non-interactive in our sense: a player can time-stamp a document while
only receiving information \cite{MST04}.  
However, no reasonable BC
or protocol that time-stamps a bit
exist  in this model. It is straightforward to see that any
such protocol can be broken by an adversary with classical memory of
size twice that of an honest player, while our protocol requires no memory
for the honest players and remains secure against any adversary not able
to store more than half the size of the quantum transmission. 

We also note that it has been shown earlier that BC is possible using
quantum communication, assuming a different type of physical
limitation, namely a bound on the size of coherent measurement that
can be implemented \cite{Salvail98}. This limitation is incomparable to
ours: it does not limit the total size of the memory, instead it
limits the number of bits that can be simultaneously operated on to
produce a classical result. Our adversary has a limit on the total
memory size, but can measure all of it coherently. The protocol from
\cite{Salvail98} is interactive, and requires a bound on the maximal
measurement size that is sub-linear in $n$.

On the technical side, we derive a new type of uncertainty relation
involving the min-entropy of a quantum encoding (Theorem~\ref{thm:mub}
and Corollary~\ref{cor:generalhmax}). The relation is in a suitable
form to apply privacy amplification against quantum adversaries as
introduced by Renner and K\"onig~\cite{RK05}.


\section{Preliminaries}
\subsection{Notation}
For a set $I=\{i_1, i_2, \ldots, i_{\ell} \} \subseteq \{1, \ldots,
n\}$ and a $n$-bit string $x \in \nbit$, we define $x|_I \assign
x_{i_1} x_{i_2} \cdots x_{i_{\ell}}$.  For $x,y \in \nbit$, $x \cdot y
\in \set{0,1}$ denotes the (standard) in-product of $x$ and~$y$.  For
a probability distribution $Q$ over $n$-bit strings and a set $L
\subseteq \nbit$, we abbreviate the (overall) probability of $L$ with
$Q(L) \assign \sum_{x \in L} Q(x)$. All logarithms in this paper are
to base two.  We denote by $h(p)$ the binary entropy function
$h(p)\assign -\big(p\cdot\log{p} + (1-p)\cdot\log{(1-p)}\big)$. We
denote by $\negl{n}$ any function of $n$ smaller than any polynomial
provided $n$ is sufficiently large.  For $x \in \nbit$, we write
$\ball{\delta n}(x)$ for the set of all $n$-bit strings at Hamming
distance at most $\delta n$ from $x$. Note that the number of elements
in $\ball{\delta n}(x)$ is the same for all $x$, we denote it by
$\ball{\delta n} \assign |\ball{\delta n}(x)|$. It is well known that
$\ball{\delta n} \leq 2^{n h(\delta)}$.

The pair $\{\ket{0},\ket{1}\}$ denotes the computational or
rectilinear or ``$+$'' basis for the $2$-dimensional complex 
Hilbert space $\mathbb{C}^2$. 
The diagonal or ``$\times$'' basis is defined as
$\{\ket{0}_\times,\ket{1}_{\times}\}$ where
$\ket{0}_{\times}=\frac{1}{\sqrt{2}}(\ket{0}+\ket{1})$ and
$\ket{1}_{\times}=\frac{1}{\sqrt{2}}(\ket{0}-\ket{1})$.   Measuring a qubit
in the $+\,$-basis (resp.\ $\times$-basis) means applying the measurement 
described by projectors
$\ket{0}\bra{0}$ and $\ket{1}\bra{1}$
(resp. projectors $\ket{0}_\times \bra{0}_\times$ 
and $\ket{1}_{\times}\bra{1}_\times$). 
When the context requires it, we write $\ket{0}_+$ and $\ket{1}_+$ instead of $\ket{0}$ respectively $\ket{1}$; and for any
$x\in\{0,1\}^n$ and $r \in\{+,\times\}$, we write
$\ket{x}_{r}=\bigotimes_{i=1}^n\ket{x_i}_{r}$.  
If we want to choose the $+$ or $\times$-basis according to the bit $b \in
\{0,1\}$, we write $\{ +, \times \}_{[b]}$.


\subsection{Quantum Probability Theory}
As basis for the security definitions and proofs of our protocols, we
are using the formalism introduced in~\cite{RK05}, which we briefly
summarize here. A {\em random state} $\rs$ is a random variable, with
distribution $P_{\rs}$, whose range is the set of density operators of
a fixed Hilbert space. The view of an observer (which is ignorant of
the value of $\rs$) is given by the quantum system described by the
density operator $[\rs] \assign \sum_{\rho} P_{\rs}(\rho) \rho$. In
general, for any event $\cal E$, we define $[\rs|{\cal E}] \assign
\sum_{\rho} P_{\rs|{\cal E}}(\rho) \rho$. If $\rs$ is dependent on
some classical random variable $X$, with joint distribution $P_{X
  \rs}$, we also write $\rho_x$ instead of $[\rs|X=x]$. Note that
$\rho_x$ is a density operator (for any fixed $x$) whereas $\rho_X$ is
again a random state. The overall quantum system is then given by
$\mbox{$[\set{X} \otimes \rs]$} = \sum_x P_X(x) \,\set{x} \otimes \rho_x$,
where $\set{x}\assign\ket{x}\bra{x}$ is the {\em state representation}
of $x$ and $\set{X}$ the corresponding random state. Obviously,
$[\set{X} \otimes \rs] = [\set{X}] \otimes [\rs]$ if and only if
$\rho_X$ is independent of $X$, where the latter in particular implies
that no information on $X$ can be learned by observing only~$\rs$.
Furthermore, if $[\set{X} \otimes \rs]$
and \mbox{$[\set{X}] \otimes [\rs]$} are $\varepsilon$-close in terms of their
trace distance $\delta(\rho,\sigma) = \frac{1}{2} \tr(|\rho-\sigma|)$,
then the real system $[\set{X} \otimes \rs]$ ``behaves'' as the ideal
system $[\set{X}] \otimes [\rs]$ except with probability
$\varepsilon$~\cite{RK05} in that for any evolution of the system no
observer can distinguish the real from the ideal one with advantage
greater than $\varepsilon$. 
Henceforth, we use $\unif$ to denote a random variable with
range $\set{0,1}$, uniformly distributed and independent of anything
else, and, as in~\cite{RK05}, we use $d(X|\rs)$ as a short hand for
$\delta\big([\set{X} \otimes \rs],[\set{\unif}]\otimes[\rs]\big)$.

We consider the notion of the classical {\em R\'enyi entropy} $H_{\alpha}(X)$ of order $\alpha$ of a random variable $X$~\cite{Renyi61}, as well as its generalization to the R\'enyi entropy $S_{\alpha}(\rho)$ of a state $\rho$~\cite{RK05}. It holds that $S_{\alpha}([\set{X}]) = H_{\alpha}(X)$ and $S_{\alpha}([\set{X}]) \leq S_{\beta}([\set{X}])$ if $\alpha \geq \beta$. The cases that are relevant for us are the classical {\em min-entropy} $H_{\infty}(X) = -\log{\left(
\max_x P_X(x)\right)}$ as well as the {\em max} and the {\em collision Von
Neumann entropy} $S_0(\rho) = \log{\left(\mbox{rank}(\rho)\right)}$ respectively $S_2(\rho) = -\log{\left( \sum_i
    \lambda_i^2\right)}$, where $\{\lambda_i\}_i$ are the eigenvalues
of $\rho$. 


\subsection{Bounded Quantum Storage and Privacy Amplification}\label{sec:pa}
All our protocols take place in the {\em bounded quantum-storage model}, which concretely
means the following: the state of  an adversarial player may consist of an arbitrary number
of qubits, and he may perform arbitrary quantum computation. 
At a certain point in time though, we say that {\em the memory bound applies},
which means that all but $q$ of the qubits are measured.
After this point, the player is again unbounded
in (quantum) memory and computing power.
We note that our results also apply to some cases where the adversary's
memory is not bounded but is noisy in certain ways, see Section~\ref{noisymem}.

An important tool we will use is universal hashing.
A class \gu{n} of hashing functions from $\{0,1\}^n$ to $\{0,1\}$ is 
called {\em \univ} if for any pair $x, y\in\{0,1\}^n$ with $x \neq y$
\[
\big|\{\hf\in\gu{n}: \hf(x)=\hf(y)\}\big| \leq \frac{|\gu{n}|}{2}.
\]
Several \univ\ classes of hashing functions are such that evaluating and 
picking a function uniformly and at random in $\gu{n}$ 
can be done efficiently \cite{WC77,WC79}. 

\begin{theorem}[\cite{RK05}]\label{thm:pa}
  Let $X$ be distributed over $\{0,1\}^n$, and let $\rs$ be a random
  state of $q$ qubits\footnote{Remember that $\rs$ can be correlated
    with $X$ in an arbitrary way. In particular, we can think of $\rs$
    as an attempt to store the $n$-bit string $X$ in $q$ qubits.}. Let
  $F$ be the random variable corresponding to the random choice (with
  uniform distribution and independent from $X$ and~$\rs$) of a member
  of a \univ\ class of hashing functions \gu{n}.  Then
\begin{align}
  d([F(X) | \set{F} \otimes \rs]) &\leq \frac{1}{2} \, 2^{-\frac{1}{2}(S_2( [ \{X\} \otimes \rs ] )-S_0( [ \rs ]) -1)} \label{moregeneral} \\
  &\leq \frac{1}{2} \, 2^{-\frac{1}{2}({H_{\infty}(X)-q-1})}.
  \label{dbound}
\end{align}
\end{theorem}
The first inequality~\eqref{moregeneral} is the original theorem
from~\cite{RK05}, and (\ref{dbound}) follows by observing that
$\mbox{$S_2( [ \{X\} \otimes \rs ] )$} \geq S_2([\{X\}]) = H_2(X) \geq
H_{\infty}(X)$. In this paper, we essentially only use this weaker version of the
theorem.

Note that if the rightmost term of (\ref{dbound}) is negligible, i.e.
say smaller than $2^{-\varepsilon n}$, then this situation is
$2^{-\varepsilon n}$-close to the ideal situation where $F(X)$ is perfectly
uniform and independent of $\rs$ and $F$. In particular, the situations
\mbox{$F(X)=0$} and $F(X)=1$ are statistically indistinguishable given $\rs$ and $F$~\cite{FG99}. 

The following lemma is a direct consequence of Theorem~\ref{thm:pa}.
In Section~\ref{sec:qbc}, this lemma will be useful for proving the
binding condition of our commitment scheme. Recall that for $X \in
\nbit$, $\ball{\delta n}(X)$ denotes the set of all $n$-bit strings at
Hamming distance at most $\delta n$ from $X$ and $\ball{\delta n}
\assign |\ball{\delta n}(X)|$ is the number of such strings.
\begin{lemma}\label{lem:guess}
Let $X$ be distributed over $\{0,1\}^n$, let $\rs$ be a random state of $q$ qubits
and let $\bg{X}$ be a guess for $X$ given $\rs$.
Then, for all $\delta< \frac{1}{2}$ it holds that 
\[ \Pr{\big[ \bg{X} \in
    \ball{\delta n}(X) \big]} \leq 2^{-\frac{1}{2}
  (H_{\infty}(X)-q-1) + \log(\ball{\delta n})}.
\]
\end{lemma}
In other words, given a quantum memory of $q$ qubits  arbitrarily 
correlated with a classical random variable 
$X$, the probability to find $\hat{X}$ at Hamming distance
at most $\delta n$ from $X$ where  
$nh(\delta)< \frac{1}{2} (H_\infty(X)-q)$ is negligible. 

\begin{proof}
  Here is a strategy to try to bias $F(X)$ when given $\bg{X}$ and
  $F\in_R \gu{n}$: Sample $X' \in_R \ball{\delta n}(\bg{X})$ and
  output $F(X')$.  Note that, using $p_{\text{succ}}$ as a short hand
  for the probability $\Pr{\big[ \bg{X} \in \ball{\delta n}(X) \big]}$
  to be bounded,
\begin{align*}
\Pr{\big[ F(X')=F(X) \big]}
&=  \frac{p_{\text{succ}}}{\ball{\delta n}} +  
  \bigg(1- \frac{p_{\text{succ}}}{\ball{\delta n}} \bigg) \frac{1}{2} \\[1ex]
&= \frac{1}{2} + \frac{p_{\text{succ}}}{2 \cdot \ball{\delta n}},
\end{align*}
where the first equality follows from the fact that if \mbox{$X'\neq X$}
then, as $\gu{n}$ is \univ, $\Pr{\left[ F(X)=F(X') \right]}=\frac{1}{2}$.
Since the probability of correctly guessing a binary $F(X)$ given
$F$ and $\rs$ is always upper bounded by
$\frac{1}{2}+d(F(X)|\set{F}\otimes\rs)$, in combination with
Theorem~\ref{thm:pa} the above results in
\begin{equation*}
\frac{1}{2} + \frac{p_{\text{succ}}}{2 \cdot \ball{\delta n}} \leq \frac{1}{2}+ 
  \frac{1}{2} 2^{-\frac{1}{2}({H_{\infty}(X)-q-1})}
\end{equation*}
and the claim follows immediately. 
\end{proof}

\section{Rabin Oblivious Transfer}
\subsection{The Definition}\label{sec:rabin-obliv-transf}
A protocol for Rabin Oblivious Transfer (ROT) between sender Alice and
receiver Bob allows for Alice to send a bit $b$ through an erasure
channel to Bob. Each transmission delivers $b$ or an erasure with
probability $\frac12$.  Intuitively, a protocol for ROT is secure if
\begin{itemize}
\item the sender Alice gets no information on whether $b$ was
                  received or not, no matter what she does, and
\item the receiver Bob gets no information about $b$ with
  probability at least~$\frac{1}{2}$, 
   no matter what he does. 
\end{itemize}
In this paper, we are considering quantum protocols for ROT. This
means that while the inputs and outputs of the honest senders are classical,
described by random variables, the protocol may contain quantum
computation and quantum communication, and the view of a dishonest
player is quantum, and is thus described by a random state.

Any such (two-party) protocol is specified by a family
$\{(\A_n,\B_n)\}_{n>0}$ of pairs of interactive quantum circuits (i.e.
interacting through a quantum channel). Each pair is indexed by a
security parameter $n>0$, where $\A_n$ and $\B_n$ denote the circuits
for sender Alice and receiver Bob, respectively.  In order to simplify
the notation, we often omit the index $n$, leaving the dependency on it
implicit.


For the formal definition of the security requirements of a ROT
protocol, let us fix the following notation. Let $B$ denote the binary
random variable describing \A's input bit $b$, and let $A$ and $B'$
denote the binary random variables describing \B's two output bits,
where the meaning is that $A$ indicates whether the bit was received
or not. Furthermore, for a dishonest sender \dA\ (respectively \dB) let
$\rs_{\dA}$ ($\rs_{\dB}$) denote the random state describing
\smash{\dA's (\dB's)} view of the protocol. Note that for a fixed
candidate protocol for ROT, and for a fixed input distribution $P_B$,
depending on whether we consider two honest \A\ and \B, a dishonest
\dA\ and an honest \B, or an honest \A\ and a dishonest \dB, the
corresponding joint distribution $P_{BAB'}$, $P_{\rs_{\dA}AB'}$
respectively $P_{B\rs_{\dB}}$ is uniquely determined.

\begin{definition}\label{def:ROT}
A two-party (quantum) protocol $(\A,\B)$ is a {\bf (statistically) secure ROT} if the following holds. 
\begin{description}
\item[Correctness:] For honest \A\ and \B\ 
$$\Pr{[B=B'|A=1]} \geq 1 - \negl{n} \, .$$
\item[Receiver-Privacy:] For any \dA\ 
$$ d(A | \rs_{\dA})  \leq \negl{n} \, .$$

\item[Sender-Privacy:] For any \dB\ there exists an event $\cal E$ with
  $P[{\cal E}] \geq \frac{1}{2} - \negl{n}$ such that
$$\delta([B\otimes\rs_{\dB}|{\cal E}],[B]\otimes[\rs_{\dB}|{\cal E}]) \leq \negl{n} \, .$$ 
\end{description}
If any of the above trace distances equals 0, 
then the corresponding property is said to hold {\bf perfectly}. 
If one of the properties only holds with respect to a restricted class
$\mathfrak{S}$ of \dA's respectively $\mathfrak{R}$ of \dB's, then this property
is said to hold and the protocol is said to be secure {\bf against}
$\mathfrak{S}$ respectively~$\mathfrak{R}$.
\end{definition}

Receiver-privacy requires that the joint quantum state is essentially the
same as when $A$ is uniformly distributed and independent of the
sender's view, and sender-privacy requires that there exists some event
which occurs with probability at least $\frac{1}{2}$ (the event that
the receiver does not receive the bit) and under which the joint
quantum state is essentially the same as when $B$ is distributed
(according to $P_B$) independently of the receiver's view.


We warn the reader that the above definition does not guarantee 
that the ROT protocol is equivalent to an ``ideal black-box implementation''
of ROT, so it does not guarantee universal composability, for instance.
One main reason for this is that, unlike the classical case~\cite{CSSW06},
receiver-privacy as we define it does not guarantee that the input bit $b$ is
determined after the execution of ROT. In other words, \dA\ is not
necessarily bound to her input.
In fact, this is not surprising, since our model  places no limitations whatsoever
on the sender. If  \dA\ was indeed bound to her input, a straightforward
reduction would allow us to
build from ROT a statistically hiding commitment scheme where the ROT 
sender is the committer. But since the sender is unbounded,
she can always break the binding property using essentially
the standard attack against unconditionally secure quantum bit
commitment~\cite{LC97,Mayers97}. 

A more rigorous definition of Oblivious Transfer is
therefore required in order to allow for composability. 
Moreover, we see from the above that satisfying such a definition will require 
some limitation to be
placed on the sender, such as a memory bound. This would, for instance, allow
using the commitment scheme we present later in this paper with the 
ROT sender in the role of committer.
These issue will be further addressed in a forthcoming paper~\cite{DFRSS06}.


\subsection{The Protocol}\label{sec:otprot}
We introduce a quantum protocol for ROT that will be shown perfectly
receiver-private (against any sender) and statistically sender-private against any quantum memory-bounded
receiver. Our protocol exhibits some similarity with quantum conjugate
coding introduced by Wiesner~\cite{Wiesner83}.

The protocol is very simple (see Figure~\ref{fig:ot}): $\A$ picks
$x\in_R\{0,1\}^n$ and sends to
$\B$ $n$ qubits in state either $\ket{x}_+$ 
or $\ket{x}_{\times}$ 
each chosen with probability~$\frac{1}{2}$.  $\B$ then measures all
received qubits either in the rectilinear or in the diagonal basis.
With probability \smash{$\frac{1}{2}$}, $\B$ picked the right basis and gets
$x$, while any \dB\ that is forced to measure part of the state (due
to a memory bound) can only have full information on $x$ in case the
$+$-basis was used {\em or} in case the $\times$-basis was used (but
not in both cases). Privacy amplification based on any \univ\ class of
hashing functions $\gu{n}$ is then used to destroy partial information. 
(In order to
avoid aborting, we specify that if a dishonest \dA\ refuses to
participate, or sends data in incorrect format, then \B\ samples its
output bits $a$ and $b'$ both at random in $\set{0,1}$.)

\begin{myfigure}{h}
 \begin{myprotocol}{\qot$(b)$}
      \item $\A$ picks $x\in_R\{0,1\}^n$, and $r\in_R\{+,\times \}$.
      \item $\A$ sends $\ket{\psi} \assign \ket{x}_r$ 
            to $\B$ (i.e. the string $x$ in basis $r$).
      \item $\B$ picks $r'\in_R\{+,\times \}$ and measures all qubits
       of $\ket{\psi}$ in basis $r'$. Let $x'\in\{0,1\}^n$ be the
       result.
      \item $\A$ announces $r$, $\hf\in_R \gu{n}$, and $e \assign b\oplus \hf(x)$.\label{bound}
      \item $\B$ outputs $a \assign 1$ and $b' \assign e\oplus \hf(x')$ if $r'=r$ and else $a \assign 0$ and $b' \assign 0$.
 \end{myprotocol}
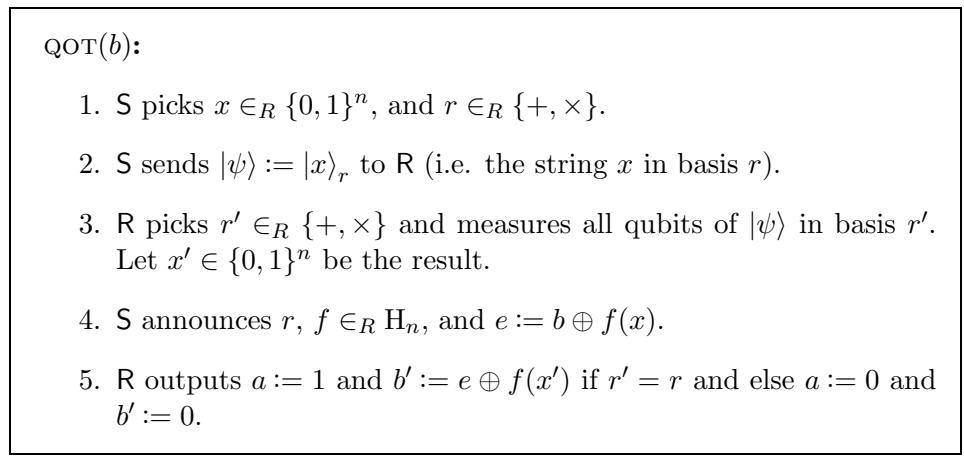
\caption{Protocol for Rabin QOT}\label{fig:ot}
\end{myfigure}

As we shall see in Section~\ref{sec:otsecurity}, the security of the
\qot\ protocol against receivers with bounded-size quantum memory holds as
long as the bound applies before Step~\ref{bound} is reached.  An
equivalent protocol is obtained by purifying the sender's actions. Although
\qot\ is easy to implement, the purified or EPR-based
version~\cite{Ekert91} depicted
in Figure~\ref{fig:eprot} is easier to prove secure. A similar
approach was taken in the Shor-Preskill proof of security for the BB84
quantum key distribution scheme \cite{SP00}.

\begin{myfigure}{h}
\begin{myprotocol}{\eprqot$(b)$}
\item $\A$ prepares $n$ EPR pairs each in state 
      $\ket{\Omega}=\frac{1}{\sqrt{2}}(\ket{00}+\ket{11})$.
\item $\A$ sends one half of each pair to $\B$ and keeps the other
  halves.\label{rec}
\item $\B$ picks $r'\in_R\{+,\times \}$ and measures all received qubits
      in basis $r'$. Let $x'\in\{0,1\}^n$ be the result.
\item $\A$ picks $r\in_R\{+,\times \}$, and measures all kept
      qubits in basis $r$. Let $x\in\{0,1\}^n$ be the outcome.  $\A$
      announces $r$, $\hf\in_R \gu{n}$, and $e \assign b\oplus \hf(x)$.\label{it:measure}
\item $\B$ outputs $a \assign 1$ and $b' \assign e \oplus \hf(x')$ if $r'=r$ and else $a \assign 0$ and $b' \assign 0$.
\end{myprotocol}
\caption{Protocol for EPR-based Rabin QOT}\label{fig:eprot}
\end{myfigure}

Notice that while \qot\ requires no quantum memory for honest players,
quantum memory for $\A$ seems to be required in \eprqot. The following
Lemma shows the strict equivalence between \qot\ and \eprqot.

\begin{lemma}\label{lem:seqequiv}
  \qot\ is secure if and only if \eprqot\ is secure.
\end{lemma}
\begin{proof}
The proof follows easily after observing that $\A$'s
choices of $r$ and $\hf$, together with 
the measurements all commute with $\B$'s actions.
Therefore, they can be performed right after Step 1
with no change for $\B$'s view. Modifying \eprqot\
that way results in \qot.
\end{proof}
Note that for a dishonest receiver it is not only irrelevant whether he tries to attack \qot\ or \eprqot, but in fact there is no difference in the two protocols from his point of view. 

\begin{lemma}\label{lem:sec:receiverprivate}
\eprqot\ is perfectly receiver-private.
\end{lemma}
\begin{proof}
  It is obvious that no information about whether
  $\B$ has received the bit is leaked to any sender \smash{$\dA$}, since $\B$ does
  not send anything, i.e. \eprqot\ is non-interactive!
\end{proof}


\subsection{Modeling Dishonest Receivers} \label{sec:modeldishonestreceivers}
We model dishonest receivers in \qot\ respectively \eprqot\ under the assumption that 
the maximum size of their quantum storage is bounded.
These adversaries are only required to have bounded quantum storage
when they reach Step \ref{bound} in (\epr-)\qot. 
Before that, 
the adversary can store and carry out quantum computations involving any number
of qubits. Apart from the restriction on the size of the quantum
memory available to the adversary, no other assumption is made. In 
particular, the adversary is not assumed to be computationally
bounded and the size of its classical memory is not restricted. 
\begin{definition}\label{boundedstorage}
The set $\mathfrak{R}_{\gamma}$ denotes all
possible quantum dishonest receivers $\{\dB_n\}_{n>0}$ in \qot\ or \eprqot\ 
where for each $n>0$, $\dB_n$ has quantum memory of size 
at most $\gamma n$ when Step \ref{bound} is reached.  
\end{definition}
In general, the adversary $\dB$ is allowed to perform any quantum
computation compressing the $n$ qubits
received from $\A$ into a quantum register $M$ of size at most $\gamma n$
when Step \ref{bound} is reached. More precisely, the compression
function is implemented by some unitary transform $C$ acting
upon the quantum state received and an ancilla of arbitrary
size. The compression is performed by a measurement that we
assume in the computational basis without loss of generality. 
Before starting Step \ref{bound}, the adversary first applies
a unitary transform $C$:
\begin{eqnarray*}
2^{-n/2}\sum_{x\in\{0,1\}^n}\ket{x}\otimes C\ket{x}\ket{0}
  \mapsto  2^{-n/2} \sum_{x\in\{0,1\}^n} \ket{x} \otimes
 \sum_{y}\alpha_{x,y}\ket{\varphi_{x,y}}^{M}\ket{y}^{Y}, 
\end{eqnarray*}
where for all $x$, $\sum_y |\alpha_{x,y}|^2=1$.
Then, a measurement in the computational basis is applied
to register $Y$ providing classical outcome $y$. The result
is a quantum state in register $M$ of size $\gamma n$ qubits.
Ignoring the value of $y$ to ease the notation,
the re-normalized state of the system is now in its most general
form when Step~\ref{bound} in \eprqot\ is reached:
\[ \ket{\psi}= 
\sum_{x\in\{0,1\}^n} \alpha_x \ket{x}\otimes\ket{\varphi_{x}}^M,
\]
where $\sum_{x} |\alpha_x|^2=1$.

\subsection{Uncertainty Relation}\label{sec:uncertainty}
We first prove a general uncertainty result and derive from that a
corollary that plays the crucial role in the security proof of
\eprqot\ and thus of \qot.
The uncertainty result concerns the situation where the sender holds an
arbitrary quantum register of $n$ qubits. He may measure them in either the
$+$ or the $\times$ basis. We are interested in the distribution of
both these measurement results, and we want to claim that they cannot
\emph{both} be ``very far from uniform''. One way to express this is to say
that a distribution is very non-uniform if one can identify a subset
of outcomes that has much higher probability than for a uniform
choice. Intuitively, the theorem below says that such sets cannot be
found for both of the sender's measurements. 
\begin{theorem} \label{thm:hadamard}
  Let the density matrix $\rho^A$ describe the state of a $n$-qubit
  register $A$.  Let $\Qp(\cdot)$ and $\Qt(\cdot)$ be the respective
  distributions of the outcome when register $A$ is measured in the
  $+$-basis respectively the $\times$-basis. Then, for any two sets
  $L^+ \subset \nbit$ and $L^{\times} \subset \nbit$ it holds that
\[ \Qp(L^+)+\Qt(L^{\times}) \leq \left(1+   \sqrt{2^{-n} |L^+|
   |L^{\times}|} \right)^2. \]
\end{theorem}
\begin{proof}
  We can purify register $A$ by adding a register $B$, such that the
  state of the composite system is pure. It can then be written as
  $\ket{\psi}^{AB} = \sum_{x\in \nbit} \alpha_x \ket{x}^A
  \ket{\varphi_x}^B$ for some complex amplitudes $\alpha_x$ and
  normalized state vectors $\ket{\varphi_x}$.

Clearly, $\Qp(x) = |\alpha_x|^2$. 
To give a more explicit form of the distribution $\Qt$, we apply the
Hadamard transformation to register~$A$:
\[ (H^{\otimes n} \otimes \I^{B}) \ket{\psi} = \!\!\!\! \sum_{z
  \in \nbit} \!\!\!\! \ket{z}\, \otimes \!\!\!\! \sum_{x \in \nbit}
  \!\!\! 2^{-\frac{n}{2}} (-1)^{x \cdot
  z} \alpha_x \ket{\varphi_x} \] 
and obtain 
$$\Qt(z) = \Bigg| \sum\limits_{x \in \nbit} \!\!\! 2^{-\frac{n}{2}} (-1)^{x
    \cdot z} \alpha_x \ket{\varphi_x} \Bigg| ^2.$$
  
  Let $\ol{L}^+$ denote the complement of $L^+$ and $p$ its
  probability $\Qp(\ol{L}^+)$. We can now split the sum in $\Qt(z)$ in
  the following way:
\begin{align*}
\Qt(z) &= \Bigg| \sum_{x \in \nbit} 2^{-\frac{n}{2}} (-1)^{x \cdot z} \alpha_x
  \ket{\varphi_x} \Bigg| ^2 \\
&=\Bigg| \sqrt{p} \sum_{x \in \ol{L}^+} 2^{-\frac{n}{2}} (-1)^{x
  \cdot z} \frac{\alpha_x}{\sqrt{p}}
  \ket{\varphi_x} + \sum_{x \in L^+} 2^{-\frac{n}{2}} (-1)^{x
  \cdot z}  \alpha_x \ket{\varphi_x}  \Bigg| ^2 \\
&=\Bigg| \sqrt{p} \cdot \zeta_z \ket{\upsilon_z} + \sum_{x \in
  L^+}  2^{-\frac{n}{2}} (-1)^{x  \cdot z}  \alpha_x \ket{\varphi_x}  \Bigg| ^2
\end{align*}
where $\ket{\upsilon_z}$ is defined as follows: For the normalized
state $\ket{\upsilon} \assign \sum_{x \in \ol{L}^+}
\frac{\alpha_x}{\sqrt{p}} \ket{x} \ket{\varphi_x}$, $\zeta_z
\ket{\upsilon_z}$ is the $z$-component of the state $H^{\otimes n}
\ket{\upsilon} = \sum_{z} \zeta_z \ket{z} \otimes \ket{\upsilon_z}$.
It therefore holds that $\sum_z |\zeta_z|^2 =1 $. 

\def\Lp{\ell^+}
\def\Lt{\ell^{\times}}

To upper-bound the amplitudes provided by the sum
over $L^+$, we notice that the amplitude is maximized when all
unit vectors $\ket{\varphi_x}$ point in the same direction
and when $(-1)^{x \cdot z}\alpha_x=|\alpha_x|$. More formally,
\begin{align}
  \Bigg| \sum_{x \in L^+} 2^{-\frac{n}{2}} (-1)^{x \cdot z} \alpha_x
  \ket{\varphi_x} \Bigg| &\leq 2^{-\frac{n}{2}} \sum_{x \in L^+}|\alpha_x| \nonumber \\
  &\leq 2^{-\frac{n}{2}} \sqrt{\big|L^+\big|} \sqrt{\sum_{x \in L^+}
    |\alpha_x|^2}\label{eq:cauchy1} \\
  & \leq 2^{-\frac{n}{2}} \sqrt{\big|L^+\big|}, \nonumber
\end{align}
where (\ref{eq:cauchy1}) is obtained from the Cauchy-Schwarz
inequality. Using $\Lp$ and $\Lt$ as shorthands for $\big|L^+\big|$ respectively $\big|L^{\times}\big|$, we conclude that
\begin{align}
  Q^{\times}(L^{\times}) &= \sum_{z \in L^{\times}} \Qt(z) \nonumber \\
  &\leq \sum_{z \in L^{\times}} \left( \left| \sqrt{p} \cdot \zeta_z \ket{\upsilon_z} \right| +
    2^{-\frac{n}{2}}  \sqrt{\Lp} \right)^2 \nonumber \\
  &\leq p \sum_{z \in L^{\times}} |\zeta_z|^2 + 2 \cdot
  2^{-\frac{n}{2}} \sqrt{\Lp} 
\sum_{z \in L^{\times}} |\zeta_z| +  \Lt \cdot 2^{-n} \Lp \nonumber \\
\label{eq:cs2} &\leq p + 2 \cdot 2^{-\frac{n}{2}} \sqrt{\Lp} 
\sqrt{\Lt \sum_{z \in L^{\times}} |\zeta_z|^2} +  2^{-n} \Lp \Lt \\
&\leq p + 2 \sqrt{ 2^{-n} \Lp \Lt} + 2^{-n} \Lp \Lt \nonumber \\[2ex] 
\label{eq:fin} &= 1 - \Qp(L^+) + 2 \sqrt{ 2^{-n} \Lp \Lt} + 2^{-n} \Lp \Lt.
\end{align}
Inequality~(\ref{eq:cs2}) follows again from Cauchy-Schwarz while
in (\ref{eq:fin}), we use the definition of $p$. The claim of the proposition
follows after re-arranging the terms.
\end{proof}

This theorem yields a meaningful bound as long as $|L^+| \cdot
|L^{\times}| < (\sqrt{2}-1)^2 \cdot 2^n$, e.g.~if $L^+$ and
$L^{\times}$ both contain less than $2^{n/2}$ elements. If for $r \in
\{+,\times \}$, $L^{r}$ contains only the $n$-bit string with the
maximal probability of $Q^{r}$, we obtain as a corollary a
slightly weaker version of a known relation (see (9) in \cite{MU88}).
\begin{corollary}\label{cor:pmax}
Let $q_{\infty}^+$ and $q_{\infty}^{\times}$ be the maximal
probabilities of the distributions $Q^+$ and $Q^{\times}$ from
above. It then holds that $q_{\infty}^+ \cdot q_{\infty}^{\times} \leq
\frac{1}{4} (1+c)^4$ where $c=2^{-n/2}$.
\end{corollary}

Theorem~\ref{thm:hadamard} can be generalized to more than two
mutually unbiased bases. We call different sets $\Ba{0}, \Ba{1},
\ldots, \Ba{N}$ of bases of the complex Hilbert space
$\mathbb{C}^{2^n}$ \emph{mutually unbiased}, if for all $i \neq j \ 
\in \{0, \ldots, N \}$, it holds that
\[ \forall \ket{\varphi} \in \Ba{i} \: \forall \ket{\psi} \in \Ba{j} : 
  \left| \braket{\varphi}{\psi} \right| ^2 = 2^{-n}.\]
\begin{theorem} \label{thm:mub}
  Let the density matrix $\rho^A$ describe the state of a $n$-qubit
  register $A$ and let $\Ba{0}, \Ba{1}, \ldots, \Ba{N}$ be mutually
  unbiased bases of register $A$. Let $Q^0(\cdot), Q^1(\cdot), \ldots,
  Q^N(\cdot)$ be the distributions of the outcome when register $A$ is
  measured in bases $\Ba{0}, \Ba{1}, \ldots, \Ba{N}$, respectively.
  Then, for any sets $L^0, L^1, \ldots, L^N \subset \nbit$, it holds that
\begin{align*}
\sum_{i=0}^N& Q^i(L^i)
\leq \: 1 - {N+1 \choose 2} + \!\!\!\!\sum_{0\leq j<k \leq N}\!\! \left(1+ \sqrt{2^{-n} |L^j| |L^k|} \right)^2.
\end{align*}
\end{theorem}
\begin{proof}
  Like in the proof of Theorem~\ref{thm:hadamard}, we can purify
  register $A$ by adding a register $B$. 
  The composite state can then be written as $\ket{\psi}^{AB} =
  \sum_{x\in \nbit} \alpha_x \ket{x}^A \ket{\varphi_x}^B$ for some
  complex amplitudes $\alpha_x$ and normalized state vectors $\ket{\varphi_x}$.

  We prove the statement by induction over $N$:
  For $N=1$, by applying an appropriate unitary transform to the whole
  system, we can assume without loss of generality that $\Ba0$ is the
  standard $+$-basis.
  
  Let us denote by $T$ the matrix of the basis change from $\Ba0$ to
  $\Ba1$. As the inner product between states $\ket{\phi} \in \Ba0$
  and $\ket{\phi'} \in \Ba1$ is always $|\braket{\phi}{\phi'}| =
  2^{-n/2}$, it follows that all entries of $T$ are complex numbers of
  the form $2^{-n/2} \cdot e^{i \lambda}$ for real $\lambda \in \RR$. 
  
  It is easy to verify that the same proof as for
  Theorem~\ref{thm:hadamard} applies after replacing the Hadamard
  transform $H^{\otimes n}$ on the sender's part  by
  $T$ and using the above observation about the entries of $T$.

\def\L#1{\ell_{#1}}
 
  For the induction step from $N$ to $N+1$, we define $p \assign Q^0(\ol{L}^0)$,
  $\ket{\upsilon} \assign \sum\limits_{x \in \ol{L}^0}
  \frac{\alpha_x}{\sqrt{p}} \ket{x} \ket{\varphi_x}$, and let $\zeta^j_z
  \ket{\upsilon^j_z}$ be the $z$-component of the state
  $\ket{\upsilon}$ transformed into basis $\Ba{j}$. As in the proof of
  Theorem~\ref{thm:hadamard}, using $\L{i}$ as a short hand for $\big| L^i \big|$, it follows:

\begin{align*}
\sum_{i=1}^N Q^i(L^i) &= \sum_{i=1}^N \sum_{z \in L^i} Q^i(z) \\
&\leq \sum_{i=1}^N \sum_{z \in L^i} \left( \sqrt{p} \left| \zeta^i_z 
\ket{\upsilon^i_z} \right| + 2^{-n/2} 
  \sqrt{\L{0}} \right)^2 \\
\begin{split}&\leq p \cdot \sum_{i=1}^N  \sum_{z \in L^i} |\zeta^i_z|^2 +
\sum_{i=1}^N \left( 
2 \cdot \sqrt{ 2^{-n} \L{0} \L{i}} + 2^{-n} \L{0} \L{i} \right) \end{split}\\
&\leq p \cdot \sum_{i=1}^N  P^i(L^i) + 
\sum_{i=1}^N \left( 1- \sqrt{ 2^{-n} \L{0} \L{i}} \right) ^2 - N 
\end{align*}
where the distributions $P^i$ are obtained by measuring register $A$
of the normalized state $\ket{\upsilon}$ in the mutually unbiased
bases $\Ba{1}, \Ba{2}, \ldots, \Ba{N}$. We apply the induction
hypothesis to the sum of $P^i(L^i)$:
\begin{align*}
  \sum_{i=1}^N Q^i(L^i) &\leq p \cdot \sum_{i=1}^N P^i(L^i) +
  \sum_{i=1}^N \left( 1+ \sqrt{ 2^{-n}
      \L{0} \L{i}} \right) ^2 - N  \\
  \begin{split}&\leq \left[1-Q^0(L^0) \right] \bigg[\sum_{1 \leq j < k
  \leq N} \!\! \Big(1+ \sqrt{2^{-n} \L{j}
        \L{k}} \Big)^2 + 1 - {N \choose 2} \bigg]\\
  &\qquad +\sum_{i=1}^N \left( 1- \sqrt{ 2^{-n} \L{0}
      \L{i}} \right) ^2 - N \end{split} \\
  \begin{split}&\leq -Q^0(L^0) + 1 - {N+1 \choose 2} +  \sum_{0 \leq j < k \leq N} \!\!
    \left(1+ \sqrt{2^{-n} \L{j} \L{k}} \right)^2\end{split}
\end{align*}
where the last inequality follows by observing that the term in the
right bracket is at least $1$ and rearranging the terms. This
completes the induction step and the proof of the proposition.
\end{proof}

Analogous to Corollary~\ref{cor:pmax}, we derive an uncertainty relation about
the sum of the min-entropies of up to $2^\frac{n}{4}$ distributions.
\begin{corollary} \label{cor:generalhmax}
  For an $\varepsilon>0$, let $0< N < 2^{(\frac{1}{4}-\varepsilon)n}$.
  For $i=0,\ldots,N$, let $H_{\infty}^i$ be the min-entropies of the
  distributions $Q^i$ from the theorem above. Then,
\[ \sum_{i=0}^N H_{\infty}^i \geq (N+1) \big(\log (N+1) - \negl{n}\big). \]
\end{corollary}
\begin{proof}
  For $i=0,\ldots,N$, we denote by $q_{\infty}^i$ the maximal
  probability of $Q^i$ and let $L^i$ be the set containing only the
  $n$-bit string $x$ with this maximal probability $q_{\infty}^i$.
  Theorem~\ref{thm:mub} together with the assumption about $N$ assures
  $\sum_{i=0}^N q_{\infty}^i \leq 1 + \negl{n}$. By the inequality of
  the geometric and arithmetic mean follows:
\begin{align*}
\sum_{i=0}^N H_{\infty}^i &= -\log{ \prod_{i=0}^N q_{\infty}^i }
 \geq - \log \left(\frac{1 + \negl{n}}{N+1} \right)^{N+1}\\
  &= (N+1) \big(\log(N+1) - \negl{n}\big).
\end{align*}
\end{proof}

\subsection{Security Against Dishonest Receivers}\label{sec:otsecurity}
In this section, we show that \eprqot\ is secure against any dishonest
receiver having access to a quantum storage device of size strictly 
smaller than half the number of qubits received at Step \ref{rec}. 

In our setting, we use Theorem~\ref{thm:hadamard} to lower-bound the
overall probability of strings with small probabilities in the
following sense. For $0 \leq \gamma + \kappa \leq 1$, define
\begin{align*}
S^+ \assign \big\{ x \in \nbit &: \Qp(x) \leq  2^{-(\gamma + \kappa )n}
\big\}\; \mbox{ and}\\
S^{\times} \assign \big\{ z \in \nbit &: \Qt(z) \leq  2^{-(\gamma
  +\kappa )n} \big\}
\end{align*} 
to be the sets of strings with small probabilities and denote by $L^+
\assign \ol{S}^+$ and $L^{\times} \assign \ol{S}^{\times}$ their
complements. (Here's the mnemonic: $S$ for the strings with
  \emph{S}mall probabilities, $L$ for \emph{L}arge.) Note that
for all $x \in L^+$, we have that $\Qp(x) > 2^{-(\gamma + \kappa )n}$
and therefore $|L^+| < 2^{(\gamma + \kappa)n}$. Analogously, we have
$|L^{\times}| < 2^{(\gamma + \kappa)n}$. For the ease of notation, we
abbreviate the probabilities that strings with small probabilities
occur as follows: $\qp \assign \Qp(S^+)$ and $\qt \assign \QtSt$. The
next corollary now immediately follows from
Theorem~\ref{thm:hadamard}.
\begin{corollary} \label{cor:hadamard}
  Let $\gamma + \kappa < \frac{1}{2}$. For the probability
  distributions $\Qp$, $\Qt$ and the sets $S^+$, $S^{\times}$ defined
  above, we have
\[ \qp + \qt = \Qp(S^+) + \QtSt \geq 1 - \negl{n}. \]
\end{corollary}

\begin{theorem}\label{thm:privacyeprqot}
  For all $\gamma < \frac{1}{2}$, \qot\ is secure against
  $\mathfrak{R}_{\gamma}$.
\end{theorem}
\begin{proof}
  After Lemmata~\ref{lem:seqequiv} and~\ref{lem:sec:receiverprivate}, it
  remains to show that \eprqot\ is sender-private against
  $\mathfrak{R}_{\gamma}$.  Since \mbox{$\gamma < \frac{1}{2}$}, we can find
  $\kappa > 0$ with \mbox{$\gamma + \kappa < \frac{1}{2}$}.  Consider a
  dishonest receiver in \eprqot\ \smash{$\dB$} with quantum
  memory of size~$\gamma n$.

  Using the notation from Section~\ref{sec:rabin-obliv-transf}, we
  show that there exists an event $\cal E$ such that $P[{\cal E}] \geq
  \frac{1}{2} - \negl{n}$ as well as
  $\delta([\set{B}\!\otimes\!\rs_{\dB}|{\cal
    E}],[\set{B}]\!\otimes\![\rs_{\dB}|{\cal E}]) \leq \negl{n}$, as
  required by the sender-privacy condition of Definition~\ref{def:ROT}.
  Let $X$ denote the random variable describing the outcome $x$ of
  \A's measurement (in basis $r$) in Step~\ref{it:measure} of \eprqot.
  We implicitly understand the distribution of $X$ to be conditioned
  on the classical outcome $y$ of the measurement \smash{$\dB$}
  performs when the memory bound applies, as described in
  Section~\ref{sec:modeldishonestreceivers}.  We define $\cal E$ to be
  the event $X \in S^r$. Note that $\cal E$ is independent of $B$ and
  thus $[B|{\cal E}] = [B]$. Furthermore, due to the uniform choice of
  $r$, and using Corollary~\ref{cor:hadamard}, $P[{\cal E}] =
  \frac{1}{2} (\qp + \qt) \geq \frac{1}{2} - \negl{n}$.

In order to show the second condition, we have to show that whenever $\cal E$ occurs, the dishonest receiver cannot distinguish the situation
where $B=0$ is sent from the one where $B=1$ is sent. 
As the bit $B$ is masked by the output of the hash function $F(X)$ in Step 4 of \eprqot\ (where the random variable $F$ represents the random choice for~$f$), this is
equivalent to distinguish between $F(X)=0$ and $F(X)=1$.
This situation is exactly suited for applying
Theorem~\ref{thm:pa}, which says that $F(X)=0$ is indistinguishable from $F(X)=1$
whenever the right-hand side of (\ref{dbound}) is negligible.
\pagebreak[3]

In the case $r=+$, we have
\begin{align}
  H_{\infty}(X|X \in S^+) &= -\log\left(\max_{x \in S^+} \frac{\Qp(x)}{\qp}\right) \nonumber\\
  &\geq -\log\left(\frac{2^{-(\gamma + \kappa)n}}{\qp}\right) = \gamma
  n + \kappa n + \log(\qp). \label{eq:Hinf}
\end{align}

If $\qp \geq 2^{-\frac{\kappa}{2} n}$ then $H_{\infty}(X|X \in S^+) \geq \gamma
n + \frac{\kappa}{2} n$ and indeed the right-hand side of (\ref{dbound}) decreases
exponentially when conditioning on $X \in S^+$. The corresponding
holds for the case $r = \times$.

Finally, if $\qp < 2^{-\frac{\kappa}{2} n}$ (or similarly \mbox{$\qt <
  2^{-\frac{\kappa}{2} n}$}) then instead of as above we define $\cal E$ as the
{\em empty event} if $r = +$ and as the event $X \in S^{\times}$ if $r
= \times$. It follows that $P[{\cal E}] = \frac12 \cdot \qt \geq
\frac12 - \negl{n}$ as well as $H_{\infty}(X|{\cal E}) = H_{\infty}(X|X \in S^{\times}) \geq \gamma n + \kappa
n + \log(\qt) \geq \gamma n + \frac{\kappa}{2} n$ (for $n$ large enough), both by
Corollary~\ref{cor:hadamard} and the bound on~$\qp$.
\end{proof}

\subsection{On the Necessity of Privacy Amplification}
In this section, we show that randomized privacy amplification seems
to be needed for protocol \qot\ to be secure.
It is tempting to believe that the sender could use the
xor $\bigoplus_i x_i$ in order to mask the bit $b$, rather than $f(x)$
for a randomly sampled $f \in \gu{n}$. This would reduce the
communication complexity as well as the number of random coins needed.
However, we argue in this section that this is not secure (against an
adversary as we model it). Indeed, somewhat surprisingly, this variant
can be broken by a dishonest receiver that has {\em no quantum memory
  at all} (but that can do coherent measurements on pairs of qubits).

Clearly, a dishonest receiver can break the modified scheme \qot\ and
learn the bit $b$ with probability $1$ if he can compute $\bigoplus_i
x_i$ with probability~$1$. Note that, using the equivalence between
\qot\ and \eprqot, $x_i$ can be understood as the outcome of the
measurement in either the $+$- or the $\times$-basis, performed by the
sender on one part of an EPR pair while the other has been handed over
to the receiver.  The following proposition shows that indeed the
receiver can learn $\bigoplus_i x_i$ by a suitable measurement of his
parts of the EPR pairs. Concretely, he measures the qubits he receives
pair-wise by a suitable measurement which allows him to learn the xor
of the two corresponding $x_i$'s, no matter what the basis is (and he
needs to store one single qubit in case $n$ is odd). This obviously
allows him to learn the xor of all $x_i$'s in all cases.

\begin{proposition}
  Consider two EPR pairs, i.e., $\ket{\psi} = \frac{1}{2} \sum_{x}
  \ket{x}^S\ket{x}^R$ where $x$ ranges over $\set{0,1}^2$.  Let $r \in
  \set{+,\times}$, and let $x_1$ and $x_2$ be the result when
  measuring the two qubits in register $S$ in basis $r$.  There exists
  a fixed measurement for register $R$ so that the outcome together
  with $r$ uniquely determines $x_1 \oplus x_2$.
\end{proposition}

\begin{proof}
The measurement that does the job is the {\em Bell measurement}, i.e., the measurement in the Bell basis $\set{\ket{\Phi^+},\ket{\Psi^+},\ket{\Phi^-},\ket{\Psi^-}}$. Recall, 
\begin{align*}
\ket{\Phi^+} &= \frac{1}{\sqrt{2}} \big(\ket{00}_+ + \ket{11}_+\big) =  \frac{1}{\sqrt{2}} \big(\ket{00}_{\times} + \ket{11}_{\times}\big) \\
\ket{\Psi^+} &= \frac{1}{\sqrt{2}} \big(\ket{01}_+ + \ket{10}_+\big) =  \frac{1}{\sqrt{2}} \big(\ket{00}_{\times} - \ket{11}_{\times}\big) \\
\ket{\Phi^-} &= \frac{1}{\sqrt{2}} \big(\ket{00}_+ - \ket{11}_+\big) =  \frac{1}{\sqrt{2}} \big(\ket{01}_{\times} + \ket{10}_{\times}\big) \\
\ket{\Psi^-} &= \frac{1}{\sqrt{2}} \big(\ket{01}_+ - \ket{10}_+\big) =  \frac{1}{\sqrt{2}} \big(\ket{10}_{\times} - \ket{01}_{\times}\big)  \, .
\end{align*}
Due to the special form of the Bell basis, when register $R$ is
measured and, as a consequence, one of the four Bell states is
observed, the state in register $S$ collapses to that {\em same} Bell
state. Indeed, when doing the basis transformation, all cross-products
cancel each other out. It now follows by inspection that knowledge of
the Bell state and the basis $r$ allows to predict the xor of the two
bits observed when measuring the Bell state in basis $r$. For
instance, for the Bell state $\ket{\Psi^+}$, the xor is $1$ if $r = +$
and it is $0$ if $r = \times$.
\end{proof}

Note that from the above proof one can see that the receiver's attack,
respectively his measurement on each pair of qubits, can be understood
as teleporting one of the two (entangled) qubits from the receiver to
the sender using the other as EPR pair (but the receiver does not send
the outcome of his measurement to the sender, but keeps it in order to
predict the xor). 

Clearly, the same strategy also works against any 
fixed linear function. Therefore, the only hope for doing deterministic privacy amplification is by using a non-linear function; but whether it is possible at all is not known to us.


\subsection{Weakening the Assumptions}\label{sec:weakass}
Observe that \qot\ requires error-free quantum communication, in that
a transmitted bit $b$, that is encoded by the sender and measured by
the receiver using the same basis, is always received as $b$. And it
requires a perfect quantum source which on request produces {\em one}
qubit in the right state, e.g.\ {\em one} photon with the right
polarization.  Indeed, in case of noisy quantum communication, an
honest receiver in \qot\ is likely to receive an incorrect bit, and
the sender-privacy of \qot\ is vulnerable to imperfect sources that once in
while transmit more than one qubit in the same state: a malicious
receiver $\dB$ can easily determine the basis $r \in \{+,\times \}$
and measure all the following qubits in the right basis.  However,
current technology only allows to approximate the behavior of
single-photon sources and of noise-free quantum communication. It
would be preferable to find a variant of \qot\ that allows to weaken
the technological requirements put upon the honest parties.

In this section, we present such a protocol based on BB84 states
\cite{BB84}, \BBqot\ (see Figure~\ref{fig:BB84ot}). The security proof
follows essentially by adapting the security analysis of \qot\ in a
rather straightforward way, as will be discussed later.

Let us consider a quantum channel with an error probability $\phi <
\frac{1}{2}$, i.e., $\phi$ denotes the probability that a transmitted
bit $b$, that is encoded by the sender and measured by the receiver
using the same basis, is received as $1-b$. In order not to have the
security rely on any level of noise, we assume 
the error probability to be zero when considering a {\em dishonest} receiver.
Also, let us consider a quantum source which produces two or more
qubits (in the same state), rather than just one, with probability
$\eta < 1 - \phi$.
We call this the $(\phi,\eta)$-weak quantum model.

In order to deal with noisy quantum communication, we need to do
error-correction without giving the adversary too much information. 
\remove{ 
For this, we use {\em secure sketches}, as introduced in~\cite{DRS04}.
A $(\ell,m,\phi)$-secure sketch\footnote{Note that our definition of a
  secure sketch differs slightly from the one given in~\cite{DRS04}. }
is a randomized function $S:\{0,1\}^{\ell} \rightarrow \{0,1\}^*$ such
that (1) for any $w \in \{0,1\}^{\ell}$ and for $w'$ received from $w$
by flipping each bit (independently) with probability $\phi$, the
string $w$ can be recovered from $w'$ and $S(w)$ except with
negligible probability (in $\ell$), and (2) for all random variables
$W$ over $\{0,1\}^{\ell}$, the ``average min-entropy'' of $W$ given
$S(W)$ is at least $H_{\infty}(W) - m$.  We would like to point out
that the notion of average min-entropy used in \cite{DRS04} and here differs
slightly from the standard notion $H_{\infty}(W|S(W))$, but it implies
that for any $\Delta > 0$, the probability that $S(W)$ takes on a
value $y$ such that $H_{\infty}(W|S(W)=y) \geq H_{\infty}(W)-m-\Delta$
is at least $1-2^{-\Delta}$ (which is sufficient for our purpose).
}
Techniques to solve this problem are known as {\em information
  reconciliation} (e.g.~\cite{BS93}) or as {\em secure
  sketches}~\cite{DRS04}.  Let $x \in \{0,1\}^\ell$ be an arbitrary
string, and let $x' \in\{0,1\}^\ell$ be the result of flipping every
bit in $x$ (independently) with probability $\phi$. It is well
known that learning the syndrome $S(x)$ of $x$, with respect to an
efficiently decodable linear error-correcting code $C$ of length
$\ell$ with minimal distance $d = (\phi+\varepsilon)\ell$ where
$\varepsilon > 0$, allows to recover $x$ from $x'$, except
with negligible probability in $\ell$ (e.g.~\cite{Maurer91,Crepeau97,DRS04}).
Furthermore, it is known from coding theory that (for large enough
$\ell$) such a code can be chosen with rate $R$ arbitrary close to
(but smaller than) $1 - h(\phi)$, i.e., such that the syndrome length
$s$ is bounded by $s < (h(\phi) + \varepsilon) \ell$ where
$\varepsilon > 0$ (see e.g.~\cite{Crepeau97} and the reference therein).

Regarding the loss of information, we can analyze privacy amplification 
in a similar way as before, just by adding the syndrome 
$S(x)$ to the random state $\rs$. Using that $S_0([\set{S(X)} \!\otimes\! \rs]) \leq q + s$, Theorem~\ref{thm:pa} then reads as
\begin{align}
  d(F(X) \,|\set{F}\!\otimes\!\set{S(X)} \!\otimes\! \rs) &\leq \frac{1}{2} 2^{-\frac{1}{2}({H_{\infty}(X)-q-s-1})}.
\end{align}
%

Consider the protocol \BBqot\ in the $(\phi,\eta)$-weak quantum model
shown in Figure~\ref{fig:BB84ot}. The protocol uses a efficiently decodable linear code $C_{\ell}$, parameterized in $\ell \in \NN$, with codeword length $\ell$, minimal distance $d = (\phi+\varepsilon)\ell$, and rate $R = 1 - h(\phi) - \varepsilon$ for some small $\varepsilon > 0$. Let $S_{\ell}$ be the corresponding syndrome function. 
Like before, the
memory bound in \BBqot\ applies before Step~\ref{BBbound}.

\begin{myfigure}{h}
\begin{myprotocol}{\BBqot$(b)$}
 \item $\A$ picks $x \in_R \nbit$ and $\theta \in_R \{+,\times \}^n$.
 \item $\A$ sends $x_i$ in the corresponding bases $\ket{x_1}_{\theta_1}, \ldots, \ket{x_n}_{\theta_n}$ to~$\B$.
\item $\B$ picks $r'\in_R\{+,\times \}$ and measures all qubits in
    basis $r'$. Let $x'\in\{0,1\}^n$ be the result.
\item $\A$ picks $r \in_R \{+,\times \}$, sets $I \assign \Set{i}{\theta_i \!=\!\set{+,\times}_{[r]}}$ and $\ell \assign |I|$, and announces $r$, $I$, $syn \assign S_{\ell}(x|_{I})$, $\hf\in_R \gu{\ell}$, and $e \assign b\oplus \hf(x|_{I})$.\label{BBbound}
\item $\B$ recovers $x|_{I}$ from $x'|_{I}$ and $syn$, and outputs
  $a \assign 1$ and $b' \assign e \oplus \hf(x|_{I})$ if $r'=r$ and else $a
  \assign 0$ and $b' \assign 0$.
\end{myprotocol}
\caption{Protocol for the BB84 version of Rabin QOT}\label{fig:BB84ot}
\end{myfigure}

By the above mentioned properties of the code $C_{\ell}$, it is obvious that $\B$ receives
the correct bit $b$ if $r'=r$, except with negligible probability.
(The error probability is negligible in $\ell$, but by Bernstein's law of large numbers, $\ell$ is linear in $n$ except with negligible probability.)
Also, since there is no communication from $\B$ to $\A$, \BBqot\ is
clearly receiver-private.  Similar as for protocol \qot, in order to argue
about sender-privacy we compare \BBqot\ with a purified version shown
in Figure~\ref{fig:BB84eprot}.  \BBeprqot\ runs in the $(\phi,0)$-weak
quantum model, and the imperfectness of the quantum source assumed in
\BBqot\ is simulated by $\A$ in \BBeprqot\ so that there is no
difference from $\B$'s point of view. 

\begin{myfigure}{h}
\begin{myprotocol}{\BBeprqot$(b)$}
\item $\A$ prepares $n$ EPR pairs each in state
  $\ket{\Omega}=\frac{1}{\sqrt{2}}(\ket{00}+\ket{11})$. Additionally, $\A$ initializes $I'_{+}:= \emptyset$
  and $I'_{\times}:= \emptyset$.
\item For every $i \in \{1,\ldots,n\}$, $\A$ does the following. With
  probability $1-\eta$ $\A$ sends one half of the $i$-th pair to $\B$
  and keeps the other half. While with probability $\eta$ $\A$ picks $\theta_i \in_R \set{+,\times}$, replaces $I'_{\theta_i}$ by $I'_{\theta_i} \cup \{i\}$ and sends two or more
  qubits in the same state $\ket{x_i}_{\theta_i}$ to $\B$ where $x_i \in_R
  \{0,1\}$.
\item $\B$ picks $r'\in_R\{+,\times \}$ and measures all received
  qubits in basis $r'$. Let $x'\in\{0,1\}^n$ be the result.
\item\label{it:abort} $\A$ picks a random index set $J \subset_R
  \{1,\ldots,n\} \setminus (I'_{+}\cup I'_{\times})$. Then, it picks $r \in_R \{+,\times \}$, sets $I \assign J \cup I'_r$ and $\ell \assign |I|$, and for each $i \in J$ it measures the corresponding qubit
  in basis $r$. Let $x_i$ be the corresponding outcome, and let
  $x|_{I}$ be the collection of all $x_i$'s with $i \in
  I$. $\A$ announces $r$, $I$, $syn =
  S_{\ell}(x|_{I})$, $\hf\in_R \gu{\ell}$, and $e = b\oplus
  \hf(x|_{I})$.
\item $\B$ recovers $x|_{I}$ from $x'|_{I}$ and
  $syn$, and outputs $a \assign 1$ and $b' \assign e \oplus \hf(x|_{I})$, if
  $r'=r$ and else $a \assign 0$ and $b' \assign 0$.
\end{myprotocol}
\caption{Protocol for EPR-based Rabin QOT, BB84 version}\label{fig:BB84eprot}
\end{myfigure}
The security equivalence between \BBqot\ (in the $(\phi,\eta)$-weak
quantum model) and \BBeprqot\ (in the $(\phi,0)$-weak quantum model)
is omitted here as it follows essentially along the same lines as in
Section~\ref{sec:otprot}. 

\begin{theorem}\label{thm:BBqotsec}
In the $(\phi,\eta)$-weak quantum model, \BBqot\ is secure against
$\mathfrak{R}_{\gamma}$ for any 
$\gamma < \frac{1-\eta}{4} - \frac{h(\phi)}{2}$ 
(if parameter $\varepsilon$ is chosen small enough).
\end{theorem}
\begin{sketch}
  It remains to show that \BBeprqot\ is sender-private against
  $\mathfrak{B}_{\gamma}$ (in the $(\phi,0)$-weak quantum model).  The
  reasoning goes exactly along the lines of the proof of
  Theorem~\ref{thm:privacyeprqot}, except that we restrict our
  attention to those $i$'s which are in $J$. 
By Bernstein's law of large numbers, $\ell$ lies within $(1\pm\varepsilon)n/2$ and $|J|$ within $(1-\eta\pm\varepsilon)n/2$ except with negligible probability. In order to make the proof easier to read, we assume that $\ell = n/2$ and $|J| = (1-\eta)n/2$, and we also treat the $\varepsilon$ occurring in the rate of the code $C_{\ell}$ as zero. For the full proof, we simply need to carry the $\varepsilon$'s along, and then choose it small enough at the end of the proof.  

Write $n' = |J| = (1-\eta)n/2$, and let $\gamma'$ be such that $\gamma n =
  \gamma' n'$, i.e., $\gamma' = 2\gamma/(1-\eta)$. 
Let $S^+$ and $S^{\times}$ be defined as in Section~\ref{sec:otsecurity}, but with respect to $n'$ and $\gamma'$ (and some $\kappa < \frac12 - \gamma'$). 
It then
  follows as in the proof of Theorem~\ref{thm:privacyeprqot} that
\begin{align*}
H_{\infty}\big(X|_J \big| X|_J \in S^+\big) &\geq \gamma' n' +
 \kappa n' + \log(\qp)\\
 &= \gamma n + \kappa (1-\eta)n/2 + \log(\qp).
\end{align*}
Similar as in the proof of Theorem~\ref{thm:privacyeprqot}, one can make a case distinction on $\qp$ (whether $\qp \geq 2^{-\varepsilon n}$ or $\qp < 2^{-\varepsilon n}$), and in both cases argue that the min-entropy in question is larger than $\gamma n + \kappa (1-\eta)n/2$ ($\pm$ some $\varepsilon n$'s). 
By~(\ref{dbound}), it remains to argue that this is larger than $q + s = \gamma n + h(\phi) n/2$, i.e., 
$$
\kappa (1-\eta) > h(\phi) \, ,
$$ 
where $\kappa$ has to satisfy 
$$
\kappa < \frac12 -
\gamma' = \frac12 - 2\gamma/(1-\eta) \, .
$$
This can obviously be achieved (by choosing $\kappa$ appropriately) if and only if
the claimed bound on $\gamma$ holds. 
\end{sketch}

\section{Quantum Commitment Scheme}\label{sec:qbc}
In this section, we present a BC scheme from a committer
$\C$ with bounded quantum memory to an unbounded receiver $\V$.  The
scheme is peculiar since in order to commit to a bit, the
committer does not send anything. During the committing stage
information only goes from $\V$ to $\C$.  The security 
analysis of the
scheme uses similar techniques as the analysis of \eprqot.

\subsection{The Protocol}
The objective of this section is to present a bounded quantum-memory
BC scheme \comm\ (see Figure~\ref{fig:comm}). Intuitively, a
commitment to a bit $b$ is made by measuring random BB84-states in
basis $\{+,\times\}_{[b]}$.
  
\begin{myfigure}{h}
\begin{myprotocol}{\comm$(b)$}
 \item $\V$ picks $x \in_R \nbit$ and $r \in_R \{+,\times \}^n$.
 \item $\V$ sends $x_i$ in the corresponding bases $\ket{x_1}_{r_1},
   \ket{x_2}_{r_2}, \ldots, \ket{x_n}_{r_n}$ to~$\C$.
 \item $\C$ commits to the bit $b$ by measuring all qubits in basis
   $\{+,\times \}_{[b]}$. Let $x' \in \nbit$ be the result.
 \item To open the commitment, $\C$ sends $b$ and $x'$ to $\V$.
 \item $\V$ verifies that $x_i= x_i'$ for those $i$ where $r_i =
   \{+,\times \}_{[b]}$. $\V$ accepts if and only if this is the case.
\end{myprotocol}
\caption{Protocol for quantum commitment}\label{fig:comm}
\end{myfigure}

As for the OT-protocol of Section~\ref{sec:otprot}, we present an
equivalent EPR-version of the protocol that is easier to analyze (see
Figure~\ref{fig:eprcomm}).
\begin{myfigure}{h}
\begin{myprotocol}{\eprcomm$(b)$}
\item $\V$ prepares $n$ EPR pairs each in state
  $\ket{\Omega}=\frac{1}{\sqrt{2}}(\ket{00}+\ket{11})$.
\item $\V$ sends one half of each pair to $\C$ and keeps the other halves.
\item $\C$ commits to the bit $b$ by measuring all received qubits in basis
  $\{+,\times \}_{[b]}$. Let $x' \in \nbit$ be the result.
\item To open the commitment, $\C$ sends $b$ and $x'$ to $\V$.\label{bound2}
\item $\V$ measures all his qubits in basis $\{+,\times \}_{[b]}$ and
  obtains $x \in \nbit$.  He chooses a random subset $I \subseteq \{1,
  \ldots ,n\}$. $\V$ verifies that
  $x_i= x_i'$ for all $i \in I$ and accepts if and only if this is the
  case.\label{step:last}
\end{myprotocol}
\caption{Protocol for EPR-based quantum commitment}\label{fig:eprcomm}
\end{myfigure}

\begin{lemma}\label{lem:commeprcomm}
  \comm\ is secure if and only if \eprcomm\ is secure.
\end{lemma}
\begin{proof}
The proof uses similar reasoning as the one for Lemma~\ref{lem:seqequiv}. 
  First, it clearly makes no difference, if we change Step~\ref{step:last} to the
  following:
\begin{itemize}
\item[\ref{step:last}'.] $\V$ chooses the subset $I$, measures all
  qubits with index in $I$ in basis $\{+,\times \}_{[b]}$ and all
  qubits not in $I$ in basis $\{+,\times \}_{[1-b]}$. $\V$ verifies
  that $x_i= x_i'$ for all $i\in I$ and accepts if and only if this is
  the case.
\end{itemize}
Finally, we can observe that the view of $\C$ does not change if $\V$
would have done his choice of $I$ and his measurement already in
Step~1. Doing the measurements at this point means that the qubits to
be sent to $\C$ collapse to a state that is distributed identically to
the state prepared in the original scheme. The EPR-version is
therefore equivalent to the original commitment scheme from $\C$'s
point of view.
\end{proof}

It is clear that \eprcomm\ is hiding, i.e., that the commit phase reveals no information on the committed bit, since no
information is transmitted to $\V$ at all. Hence we have
\vspace{-2mm}
\begin{lemma}\label{lem:sec:hiding}
\eprcomm\ is perfectly hiding.
\end{lemma}

\subsection{Modeling Dishonest Committers}\label{sec:dishonest Comm}
A dishonest committer $\dC$ with bounded memory of at most $\gamma n$
qubits in \eprcomm\ can be modeled very similarly to the dishonest
OT-receiver $\dB$ from Section~\ref{sec:modeldishonestreceivers}:
$\dC$ consists first of a circuit acting on all $n$ qubits
received, then of a measurement of all but at most $\gamma n$ qubits,
and finally of a circuit that takes the following input: a bit $b$ that
$\dC$ will attempt to open, the $\gamma n$ qubits in memory, and some
ancilla in a fixed state. The output is a string $x' \in \nbit$ to be
sent to $\V$ at the opening stage.
\begin{definition}
  We define $\mathfrak{C}_{\gamma}$ to be the class of all committers
  $\{\dC_n\}_{n>0}$ in \comm\ or \eprcomm\ that, at the start of the opening phase
  (i.e. at Step \ref{bound2}), have a quantum memory of size at most
  $\gamma n$ qubits.
\end{definition}
We adopt the binding condition for quantum BC from \cite{DMS00}:
\begin{definition} \label{bindef}
A (quantum) BC 
scheme is {\bf (statistically) binding} against $\mathfrak{C}$ if for all
$\{\dC_n\}_{n>0}\in \mathfrak{C}$,
the probability $p_b(n)$ that $\dC_n$ 
opens $b\in\{0,1\}$ with success satisfies 
\[ p_0(n)+p_1(n) \leq 1+\negl{n}.
\] 
\end{definition}
In the next section, we show that \eprcomm\ is binding against
$\mathfrak{C}_{\gamma}$ for any $\gamma<\frac{1}{2}$. 

Note that the binding condition given here in Definition~\ref{bindef}
is weaker than the classical one, where one would require that a bit
$b$ exists such that $p_b(n)$ is negligible.  In the context of {\em
  quantum} bit commitment, this weaker definition is typically
justified by the argument that this is the best that can be achieved
for a general quantum adversary who can always commit to 0 and 1 in
superposition.  However, an adversary with bounded quantum storage
cannot necessarily maintain a commitment in superposition since the
memory compression may force a collapse.  Indeed, in upcoming work, we
show that commitment schemes exist satisfying the stronger binding
condition in the bounded quantum-storage model \cite{DFRSS06}.
While the weaker condition is sufficient for many applications, the
stronger one seems to be necessary in some cases (see the conclusion).

 

\subsection{Security Proof of the Commitment Scheme}
Note that the first three steps of \eprqot\ and
\eprcomm\ (i.e.~before the memory bound applies) are exactly the same!
This allows us to reuse Corollary~\ref{cor:hadamard} and the analysis of
Section~\ref{sec:otsecurity} to prove the binding property of
\eprcomm. 
\begin{theorem}\label{bind}
For any $\gamma<\frac{1}{2}$,
\comm\ is perfectly hiding and statistically binding against $\mathfrak{C}_{\gamma}$.  
\end{theorem}
The proof is given below. It boils down to showing that essentially
$p_0(n) \leq 1 - \qp$ and $p_1(n) \leq 1 - \qt$. The binding property
then follows immediately from Corollary~\ref{cor:hadamard}. The
intuition behind $p_0(n) \leq 1 - \qp = 1 - \Qp(S^+)$ is that a committer has only a
fair chance in opening to $0$ if $x$ measured in $+$ basis has a large
probability, i.e., $x \not\in S^+$. The following proof makes this
intuition precise by choosing the $\varepsilon$ and $\delta$'s
correctly.

\begin{proof}
  It remains to show that \eprcomm\ is binding against
  $\mathfrak{C}_{\gamma}$.  Let $\kappa >0$ be such that $\gamma +
  \kappa < \frac{1}{2}$. For the parameters $\kappa$ and $\gamma$
  considered here, define $Q^+$, $S^+$ and $\qp$ as well as
  $Q^{\times}$, $S^{\times}$ and $\qt$ as in
  Section~\ref{sec:otsecurity}.  Furthermore, let $0< \delta <\frac12$ be
  such that $h(\delta) < \kappa/2$, where $h$ is the binary entropy
  function, and choose $\varepsilon > 0$ small enough such that
  $h(\delta) < (\kappa-\varepsilon)/2$. This guarantees that
  $\ball{\delta n} \leq 2^{(\kappa-\varepsilon) n/2}$ for all
  (sufficiently large)~$n$.  For every $n$ we distinguish between the
  following two cases. If $\qp \geq 2^{-\varepsilon n/2}$ then 
$$
H_\infty(X|X \in S^+) \geq
\gamma n + \kappa n + \log(\qp) \geq \gamma n + \Big(\kappa - \frac{\varepsilon}{2}\Big) n 
$$
where the first inequality is argued as in (\ref{eq:Hinf}).
Applying Lemma~\ref{lem:guess}, it follows that any guess $\bg{X}$ for
$X$ satisfies
\begin{align*}
\Pr{}\big[ \bg{X} \in \ball{\delta n}(X) \,|\, X \in S^{+} \big]
 &\leq 2^{-\frac{1}{2} (H_{\infty}(X|X \in S^+)-\gamma n-1) + \log(\ball{\delta
 n}) } \leq 2^{-\frac{\varepsilon}{4}n + \frac{1}{2}}.
\end{align*}
However, if $\bg{X} \not\in \ball{\delta n}(X)$ then sampling a random subset of the
positions will detect an error except with probability not bigger than
$2^{-\delta n}$. Hence, 
\begin{align*} 
p_0(n) &= (1-\qp) \cdot p_{0|X \not\in S^+} + \qp \cdot p_{0|X \in S^+} \\
  &\leq 1-\qp + \qp \cdot \big(2^{-\delta n}(1-2^{-\frac{\varepsilon}{4}n +
    \frac{1}{2}})+2^{-\frac{\varepsilon}{4}n + \frac{1}{2}}\big).
\end{align*}
If on the other hand $\qp < 2^{-\varepsilon n/2}$ then trivially
$$
p_0(n) \leq 1 = 1 - \qp + \qp < 1-\qp + 2^{-\varepsilon n/2}.
$$
In any case we have $p_0(n) \leq  1-\qp + \negl{n}$. 

Analogously, we derive $p_1(n) \leq 1-\qt + \negl{n}$ and conclude that
\begin{align}
p_0(n) + p_1(n) &\leq  2-\qp-\qt + \negl{n} \leq  1+ \negl{n} \label{fl1},
\end{align}
where (\ref{fl1}) is obtained from Corollary~\ref{cor:hadamard}. 
\end{proof}

\subsection{Weakening the Assumptions}
As argued earlier, assuming that a party can produce single qubits
(with probability~1) is not reasonable given current technology. Also
the assumption that there is no noise on the quantum channel is
impractical. 
It can be shown that a straightforward modification
of \comm\ remains secure in the $(\phi,\eta)$-weak
quantum model as introduced in Section~\ref{sec:weakass},
with $\phi < \frac{1}{2}$ and
$\eta < 1- \phi$.

Let \comm'\ be the modification of \comm\ where in
Step~\ref{step:last} $\V$ accepts if and only if $x_i = x'_i$ for all
{\em but about a $\phi$-fraction} of the $i$ where $r_i = \{+,\times
\}_{[b]}$. More precisely, for all but a
$(\phi+\varepsilon)$-fraction, where $\varepsilon > 0$ is sufficiently
small.
\begin{theorem}\label{thm:weakcommsec}
  In the $(\phi,\eta)$-weak quantum model, \comm'\ is perfectly hiding
  and it is binding against $\mathfrak{C}_{\gamma}$ for any $\gamma$
  satisfying $\gamma < \frac{1}{2}(1-\eta) - 2 h(\phi)$.
\end{theorem}
\begin{sketch} 
  Using Bernstein's law of large numbers, one can argue that for {\em
    honest} $\C$ and $\V$, the opening of a commitment is accepted
  except with negligible probability.  The hiding property holds using
  the same reasoning as in Lemma~\ref{lem:sec:hiding}. And the binding
  property can be argued essentially along the lines of
  Theorem~\ref{bind}, with the following modifications. Let $J$ denote
  the set of indices $i$ where $\V$ succeeds in sending a single
  qubit. We restrict the analysis to those $i$'s which are in $J$. By
  Bernstein's law of large numbers, the cardinality of $J$ is about
  $(1-\eta)n$ (meaning within $(1-\eta\pm \varepsilon)n$), except with
  negligible probability.  Thus, restricting to these $i$'s has the
  same effect as replacing $\gamma$ by $\gamma/(1-\eta)$ (neglecting
  the $\pm \varepsilon$ to simplify notation). Assuming that $\dC$
  knows every $x_i$ for $i \not\in J$, for all $x_i$'s with $i \in J$
  he has to be able to guess all but about a $\phi/(1-\eta)$-fraction
  correctly, in order to be successful in the opening. However, $\dC$
  succeeds with only negligible probability if
$$
\phi/(1-\eta) < \delta \, .$$ 
Additionally, $\delta$ must be such that 
$$
h(\delta) < \frac{\kappa}{2} \qquad\mbox{with}\qquad \frac{\gamma}{1-\eta} + \kappa < \frac{1}{2} \, .
$$ 
Both restrictions on $\delta$ hold (respectively can be achieved by choosing $\kappa$ appropriately) if 
$$
2 \, h\!\left( \frac{\phi}{1-\eta} \right) + \frac{\gamma}{1-\eta} < \frac{1}{2} \, .
$$
Using the fact that $h(\nu p) \leq \nu h(p)$ for any $\nu \geq 1$
and $0 \leq p \leq \frac12$ such that $\nu p \leq 1$, this is clearly
satisfied if
$2 h(\phi) + \gamma < \frac{1}{2}(1-\eta)$. This  
proves the claim.
\end{sketch}

\section{Generalizing the Memory Model} \label{noisymem}
The bounded quantum-storage model limits the number of physical qubits
the adversary's memory can contain. A more realistic model would
rather address the noise process the adversary's memory undergoes.
For instance, it is not hard to build a very large, but unreliable
memory device containing a large number of qubits. It is reasonable to
expect that our protocols remain secure also in a scenario where the
adversary's memory is of arbitrary size, but where some quantum
operation (modeling noise) is applied to it.
Inequality~\eqref{moregeneral} of the Privacy Amplification
Theorem~\ref{thm:pa} allows us to apply our constructions to slightly
more general memory models. In particular, all our protocols that are secure
against adversaries with memory of no more than $\gamma n$ qubits are also secure against any noise model that reduces the rank of the
mixed state $[\rs]$, held by the adversary, to at most $2^{\gamma n}$. 

An example of a noise process resulting in a reduction of $S_0([\rs])$
is an erasure channel. Assuming the $n$ initial qubits are each erased
with probability larger than $1-\gamma$ when the memory bound applies,
it holds except with negligible probability in $n$ that
$S_0([\rs])<\gamma n$.  The same applies if the noise process is
modelled by a depolarizing channel with error probability
$p=1-\gamma$. Such a depolarizing channel replaces each qubit by a
random one with probability $p$ and does nothing with probability
$1-p$.

The technique we have developped does not allow to deal with depolarizing channels
with $p < 1-\gamma$ although one would expect that some $0< p < 1-\gamma$
should be sufficient to ensure privacy against such adversaries.
The reason being that not knowing the positions where the errors occured
should make it more difficult for the adversary than when the noise process
is modelled by an erasure channel. However, it seems that our uncertainty 
relations (i.e. Theorems~\ref{thm:hadamard} and \ref{thm:mub}) are not
strong enough to address this case. Generalizing the bounded quantum-storage
model to more realistic noisy-memory models is an interesting open question.

\section{Conclusion And Further Research}
We have shown how to construct ROT and BC securely in the bounded
quantum-storage model. Our protocols require no quantum memory for 
honest players and remain secure provided the adversary has only
access to quantum memory of size bounded by a large fraction of all
qubits transmitted.  
Such a gap between the amount of storage required for
honest players and adversaries is not achievable by classical
means.
All our protocols are non-interactive and can be
implemented using current technology. 

In this paper, we only considered ROT of one bit per invocation. Our
technique can easily be extended to deal with string ROT, essentially
by using a class of two-universal functions with range
$\set{0,1}^{\ell n}$ rather than $\set{0,1}$, for some $\ell$ with
$\gamma + \ell < \frac{1}{2}$ (respectively
$<\frac{1-\eta}{4}-\frac{h(\phi)}{2}$ for \BBqot).

Although other flavors of OTs can be constructed from ROT using
standard reductions, a more direct approach would give a better ratio
between storage-bound and communication-complexity.  More general
security definitions allowing for better composition (such as
universal composability) briefly discussed at the end of
Section~\ref{sec:rabin-obliv-transf} also disserve to be studied.
Recent extensions have shown that a 1-2 OT protocol built
along the lines of \BBqot\ is secure against adversaries with bounded
quantum memory~\cite{DFRSS06}. Interestingly, the techniques used
are quite different from the ones of this paper (which appear to fail
in case of 1-2 OT), and they additionally allow to analyse and prove
secure the bit commitment scheme \comm\ with respect to the stronger
security definition, as discussed in Section~\ref{sec:dishonest Comm}.
%

\comm\ can easily be transformed into a {\em string} commitment scheme
simply by committing bitwise, at the cost of a corresponding blow-up
of the communication complexity. In order to prove this string
commitment secure, though, it is necessary that \comm\ is secure with
respect to the stronger security definition.

How to construct and in particular prove secure a more efficient
string commitment scheme is still an open problem. Furthermore, it is
still unsolved how to construct and prove secure a 1-$m$ OT protocol, 
more efficient than via the general reduction.

Finally, finding protocols secure against adversaries in more general
noisy-memory models, quickly discussed in Section~\ref{noisymem},
would certainly be a natural extension of this work to more practical settings.



%

\section*{Acknowledgements}We would like to thank the anonymous referees
for useful comments. The authors are also grateful to Renato Renner
for enlightening discussions and Charles H. Bennett for comments on
earlier drafts.
\pagebreak[4]

\bibliographystyle{chrisabbrv}
\bibliography{qip,crypto,procs}

\end{document}